\documentclass[twocolumn,showpacs,preprintnumbers,amsmath,amssymb]{revtex4}
\usepackage{amsmath}

\usepackage{mathrsfs}
\usepackage{txfonts}
\usepackage{subfigure}
\usepackage{graphicx}
\usepackage{dcolumn}
\usepackage{bm}
\usepackage{slashed}
\usepackage{color} 
\usepackage[latin9]{inputenc}
\usepackage[T1]{fontenc}
\usepackage[english]{babel}
\usepackage{soul}
\usepackage{rotating}
\usepackage{overpic}

 {\catcode`\|=\active \gdef|{\egroup\,\vrule\,\bgroup}}

\def\be{\begin{eqnarray}}
\def\ee{\end{eqnarray}}
\begin{document}


\title{Cooper minimum of high-order harmonic spectra from MgO crystal in an ultrashort laser pulse}

\author{Yiting Zhao$^{1,2}$}
\author{Xiaoqin Xu$^{1}$}
\author{Shicheng Jiang$^{3}$} 
\author{Xi Zhao$^{4}$}
\author{Jigen Chen$^{1}$} \thanks{kiddchen@126.com}
\author{Yujun Yang$^{2}$}\thanks{yangyj@jlu.edu.cn}
\affiliation{
$^1$Zhejiang Provincial Key Laboratory for Cutting Tools ,Taizhou University, Taizhou 31800, China\\
$^2$Institute of Atomic and Molecular Physics, Jilin University, Changchun 130012, China\\
$^3$State Key Laboratory of Precision Spectroscopy, East China Normal university, Shanghai 200062, China\\
$^4$J. R. Macdonald Laboratory, Department of Physics, Kansas State University, Manhattan, Kansas 66506, USA
}

\date{\today}

\begin{abstract}
{
Cooper minimum structure of high-order harmonic spectra from atoms or molecules has been extensively studied. In this paper, we demonstrate that the crystal harmonic spectra from an ultrashort  mid-infrared laser pulse also exhibit the Cooper minimum characteristic. Based on the accurate band dispersion and k-dependent transition dipole moment (TDM) from the first-principle calculations, it can be found that the harmonic spectra from MgO crystal along  $\Gamma$-X direction present a dip structure in the plateau, which is originated from the valley of TDM by examining the distribution of the harmonic intensity at the k-space. The Cooper minimum feature in crystal HHG will pave a new way to retrieve the band information of solid materials by using HHG from the ultrashort mid-infrared laser pulse.
 }
\end{abstract} \vskip 0.5in


 \vskip 1in
\maketitle
\section{Introduction}

Atoms and molecules irradiated by an intense laser pulse can produce high-order harmonic generation (HHG) \cite{Chen1,Chen2,Tzallas3,Sansone4,Goulielmakis5,Zhang6,Bian7,Du8,Yuan9,Zhu10,Lin11,Zhang12,Li13}. Under the influence of the strong laser field, an electron can be ionized from the bound state, and accelerated in the continuum state, finally it can recombine with the ion and gives rise to the emission of extreme ultraviolet (XUV) radiation \cite{Zhu10,Lan14}. The emitted XUV radiation is closely related to the photorecombination, thus it could encode the structural information on the irradiated target and can be used to study structural features of the target and in particular the Cooper minimum \cite{Farrell15,Cloux16,Schoun17}, which corresponds to the nodal structure in the bound-free transition matrix element.

Because of clear signatures, the photoionization spectroscopy has been traditionally used to observe the Cooper minimum. In the process of HHG, the photorecombination is essentially the time inverse of photoionization, therefore the Cooper minimum should also appear in the harmonic spectra from atoms or molecules. The minimum structure in HHG from atoms or aligned molecules have been extensively investigated in many works \cite{Higuet19,Higuet20,Wong21,Farrell15,Cloux16,Schoun17,Farrell18}. Recently, HHG from solids has attracted great interest because of significant applications in attocecond pulse generation and all-optical reconstruction of the band dispersion of solids \cite{Vampa22,Venkataraman23}. It has demonstrated theoretically and experimentally that the interband polarization dominates the harmonics above the band gap for MgO, ZnO, and GaAs driven by a mid-infrared laser pulse \cite{Ghimire24,Vampa25,Wu26,Vampa27}. The interband process for solid HHG can be understood by the semiclasscial recollision model \cite{Ghimire24,Vampa25,Vampa28,Lewenstein29,S. A.30,Dimitrovski31,Vampa32,Jiang33}: the electron firstly tunneling excitation from the valence band, then the acceleration on the conduction band, finally the electron-hole recombination results in the harmonic photon. Since the harmonic generation from the interband current depends strongly on the transition dipole moment (TDM) \cite{Wismer34,Chao35,Jiang36} of the solid, if there exists a zero in the matrix element between the valence band and the conduction one, analogous to the harmonic spectra from gaseous media, the Cooper minimum structure is expected to appear in the harmonic spectra from solids.

In this paper, based on the accurate band dispersion and k-dependent TDM from the first-principle calculations, we study the feature of HHG from MgO crystal in an ultrashort mid-infrared laser pulse. It is found that, the harmonic spectra from TDM by the first-principle theory show a clear dip structure, which almost does not depend on the parameters of the driving laser pulse. Through analyzing the distribution of the harmonic intensity at different crystal momentums, it is clarified that the minimum of TDM leads to the Cooper minimum structure of HHG spectra.

\section{Theory and Models}
\subsection{Semiconductor Bloch equations}
Based on the solution of two-band semiconductor Bloch equations (SBEs) \cite{Tamaya37,Fldi38,Golde39,Golde40}, we investigate the interaction of an ultrashort mid-infrared laser pulse with MgO crystal. Atomic units are used throughout this article, unless stated otherwise. A linearly polarized laser field is propagated along the $\Gamma$-X direction of MgO, and the corresponding SBEs \cite{Jiang41,Luu42,Yu43,Wu26} can be read
\begin{eqnarray}
\begin{aligned}
\frac{\partial p_{cv}\left ( \mathbf{k}, t \right )}{\partial t}= &-i\left ( E_{c}\left ( \mathbf{k} \right )-E_{v}\left ( \mathbf{k} \right )-i/T_{2}\right )p_{cv}\left ( \mathbf{k},t \right )&\\&+ i \left[ \rho_{c} \left ( \mathbf{k}, t \right ) -\rho _{v}\left ( \mathbf{k}, t \right ) \right ]  \mathbf{F}\left ( t \right )\cdot \mathbf{D}_{cv}\left ( \mathbf{k} \right ) &\\&+\mathbf{F}\left ( t \right )\cdot \bigtriangledown _{\mathbf{k}}  p_{cv}\left ( \mathbf{k}, t \right )&
\end{aligned}
\label{eq1} 
\end{eqnarray}
\begin{eqnarray}
\begin{aligned}
\frac{\partial \rho_{v}\left ( \mathbf{k},t \right )}{\partial t}= &-2\textrm{Im}\left [ \mathbf{F}\left ( t \right )\cdot\mathbf{D}_{cv}\left ( \mathbf{k} \right )p_{cv}\left ( \mathbf{k},t \right )  \right ]&\\&+\mathbf{F}\left ( t \right )\cdot \bigtriangledown _\mathbf{k}\rho _{v}\left ( \mathbf{k},t \right )&
\end{aligned}
\label{eq2}
\end{eqnarray}
\begin{eqnarray}
\begin{aligned}
\frac{\partial \rho_{c}\left ( \mathbf{k},t \right )}{\partial t}= &2\textrm{Im}\left [ \mathbf{F}\left ( t \right )\cdot\mathbf{D}_{cv}\left ( \mathbf{k} \right )p_{cv}\left ( \mathbf{k},t \right )  \right ]&\\&+\mathbf{F}\left ( t \right )\cdot \bigtriangledown _\mathbf{k}\rho _{c}\left ( \mathbf{k},t \right ) &
\end{aligned}
\label{eq3}
\end{eqnarray}
Here, $E_{v}\left( \mathbf{k} \right )\left( E_{c}\left( \mathbf{k} \right ) \right )$  is the dispersion of the highest valence (lowest conduction) band contributing to HHG , $\rho_{v}\left ( \mathbf{k},t \right )\left (\rho_{c}\left( \mathbf{k},t \right ) \right )$ is the population in the corresponding band, $p_{cv}\left ( \mathbf{k},t \right )$ and $\mathbf{D}_{cv}\left ( \mathbf{k} \right )$are the microscopic interband polarization and the transition dipole moment between the conduction and valence bands, respectively. $\mathbf{F}\left ( t \right )=\hat{\varepsilon }F\left ( t \right )$  is the laser electric field with a Gaussian envelope and $\hat{\varepsilon }$  being the polarization direction. T$_{2}$ is the interband dephasing time. In this paper, T$_{2}$ is set to a quarter-cycle of the driving laser field. 

The intraband current $\mathbf{J}_{intra}\left ( t \right )$  because of the motions of the carriers in the bands under a laser pulse is given by
\begin{eqnarray}
\mathbf{J}_{intra}\left ( t \right ) = \sum_{\lambda =c,v}\int_{BZ}\mathbf{v}_{\lambda }\left ( \mathbf{k} \right )\rho _{\lambda }\left ( \mathbf{k} ,t\right )d\mathbf{k}   \label{eqn:4}
\end{eqnarray}
where $\mathbf{v}_{\lambda }\left ( \mathbf{k} \right )=\triangledown _{\mathbf{k}}E_{\lambda }\left ( \mathbf{k} \right )$ is the group velocity and $\lambda$ is the band index. The interband polarization  $\mathbf{J}_{inter}\left ( t \right )$  from the recombination of the electron with the hole can be expressed by

\begin{eqnarray}
\mathbf{J}_{inter}\left ( t \right ) = \frac{\partial }{\partial t}\int_{BZ}\mathbf{D}_{cv}\left ( \mathbf{k} \right )p_{cv}\left ( \mathbf{k},t \right )d\mathbf{k}+c.c.  \label{eqn:5}
\end{eqnarray}
In this work we are interested in the harmonic spectrum, which is proportional to the absolute square of the projection of the Fourier-transformed total current onto the laser polarization direction,
\begin{eqnarray}
S_{HHG}\propto \left | \int_{-\infty }^{\infty} \left [ \mathbf{J}_{intra}+ \mathbf{J}_{inter} \right ] e^{i\omega t} dt \right | ^{2}  \label{eqn:6}
\end{eqnarray}

\subsection{Band structure and transition dipole moment}
The interband and intraband currents depend significantly on the band structure and TDM \cite{Ghimire24,Chao35}. By using the Vienna Abinitio Simulation Package (VASP) code \cite{Korbman44,John45}, accurate k-dependent energy bands and TDM are achieved here. Geometric optimizations of MgO crystal with symmetry group $Fm3m$ were performed within generalized gradient approximation (GGA) in the parametrization of Perdew-Burke-Ernzerhof (PBE). The energy cutoff was set to be 400 eV, and a k-point Monkhorst pack mesh of 10 $\times$10$\times$10 was used in the Brillouin zone for electronic structure calculations. The band dispersion and TDM of MgO along the $\Gamma$-X direction were calculated by the HSE06 hybrid function with parameter 
AEXX$=$0.43,  which predict a band gap of 7.8 eV, consistent with the experimental value \cite{You46}. Based on the accurate band structure as shown in Fig. 1(a), the k-dependent TDM can be given by the following two methods, where the TDM between the valence band and the conduction one from the first-principle theory can be expressed by
\begin{eqnarray}
D_{cv}\left ( \mathbf{k} \right )=\frac{i\left \langle \Phi_{c}\left ( \mathbf{k} \right ) \left | \mathbf{p} \right |\Phi_{v}\left ( \mathbf{k} \right )\right \rangle}{\left [ E_{c}\left ( \mathbf{k} \right )-E_{v}\left ( \mathbf{k} \right ) \right ]} \label{eqn:7}
\end{eqnarray}
Because MgO crystal has inversion symmetry, the TDM between the lowest conduction band and the highest valence band is a real and even function \cite{Jiang41}, as presented by the red solid line in Fig. 1(b). 

In most of previous works about solid HHG, the TDM was calculated by the first-order $\mathbf{k}\cdot\mathbf{p}$ theory \cite{Chao35}, 
\begin{eqnarray}
D_{cv}\left ( \mathbf{k} \right )=\frac{id_{0}\left [ \varepsilon _{c}\left ( 0 \right )-\varepsilon _{v}\left ( 0 \right ) \right ]}{\left [ E_{c}\left ( \mathbf{k} \right )-E_{v}\left ( \mathbf{k} \right ) \right ]}  \label{eqn:8}
\end{eqnarray}
which is valid when the carriers in conduction or valence bands are mostly populated at $\Gamma$. However, under the interaction of a strong laser pulse, electrons (holes) may travel through the entire Brillouin zone, thus the TDM from the first-order $\mathbf{k}\cdot\mathbf{p}$ approximation is not applicable to this situation. For comparison, the TDM from the first-order $\mathbf{k}\cdot\mathbf{p}$ theory is also shown by the blue dot dashed line in Fig. 1(b). Obviously, TDM in the first-principle case exists minima at k=$\pm$0.6 $\pi/a$ ($a$=4.213 \AA ) and valley structures. Yu $et$ $al$. proved that, the shape of the k-dependent TDM plays an important role in harmonic generation \cite{Chao35}. Therefore, we will discuss how the difference in shapes of TDMs effects the HHG spectra. 

\renewcommand{\figurename}{Fig.}
\begin{figure} [hbt!]
	\centering
	\subfigure{
		\begin{minipage}{4cm}
			\centering
			\includegraphics[width=5cm]{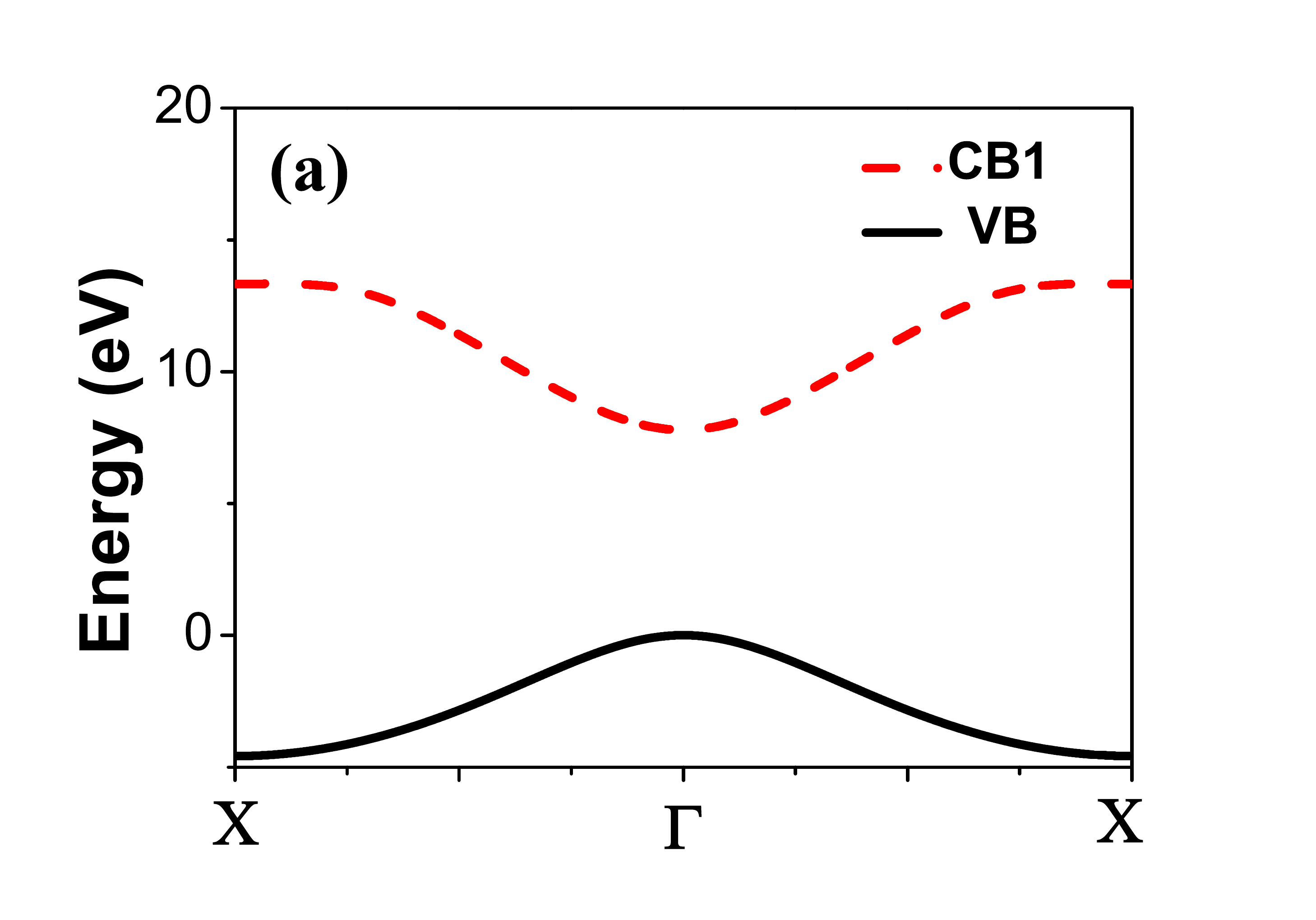}
		\end{minipage}
	}
	\subfigure{
		\begin{minipage}{4cm}
			\centering
			\includegraphics[width=5cm]{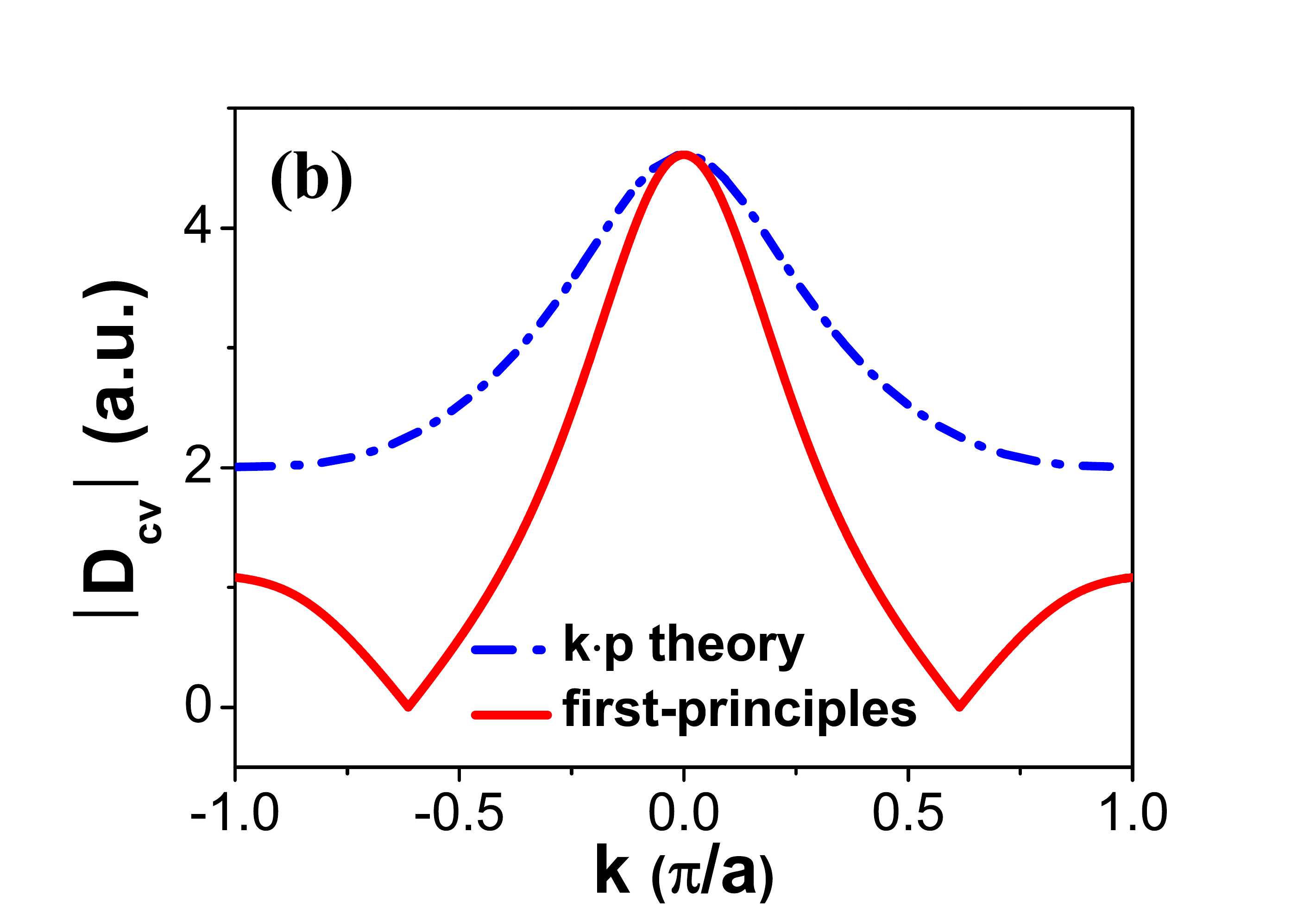}
		\end{minipage}
	}
	\caption{(a) The lowest conduction band and highest valence band of MgO along the $\Gamma$-X direction; (b) k-dependent dipole moments from the first-principle calculations (red solid curve) and first-order $\mathbf{k}\cdot \mathbf{p}$ theory (blue dot dashed curve).}
\end{figure}

\section{COOPER MINIMUM STRUCTURE OF CRYSTAL HHG}
 In terms of two-band and three-band SBEs, we firstly examine the dependence of crystal HHG spectra from the first-principle calculations on the driving laser intensity, as shown in Figs. 2(a) and 2(b), respectively. Here, the 3 fs$/$1600 nm mid-infrared laser pulse with carrier envelope phase (CEP) 0 is chosen, and the corresponding peak intensity inside the crystal is changed from $5.0\times 10^{12}$ $W/cm^{2}$ to $4.0\times 10^{13}$ $W/cm^{2}$. One can see that, as the increase of the laser intensity, HHG spectra in both cases are almost same and the harmonic efficiency is gradually enhanced; when the peak intensity of the laser pulse is stronger than $1.5\times 10^{13}$ $W/cm^{2}$, the harmonic spectra from the three-band model appear a clear two-plateau structure. In particular, harmonic spectra from the two cases exhibit an apparent minimum near 16 eV in the first plateau. In the following, for better explaining the origin of the dip structure, we focus on harmonic spectra from the two-band SBEs. 
 
 \begin{figure} [hbt!]
 	\centering
 	\subfigure{
 		\begin{minipage}{4cm}
 			\centering
 			\includegraphics[width=5cm]{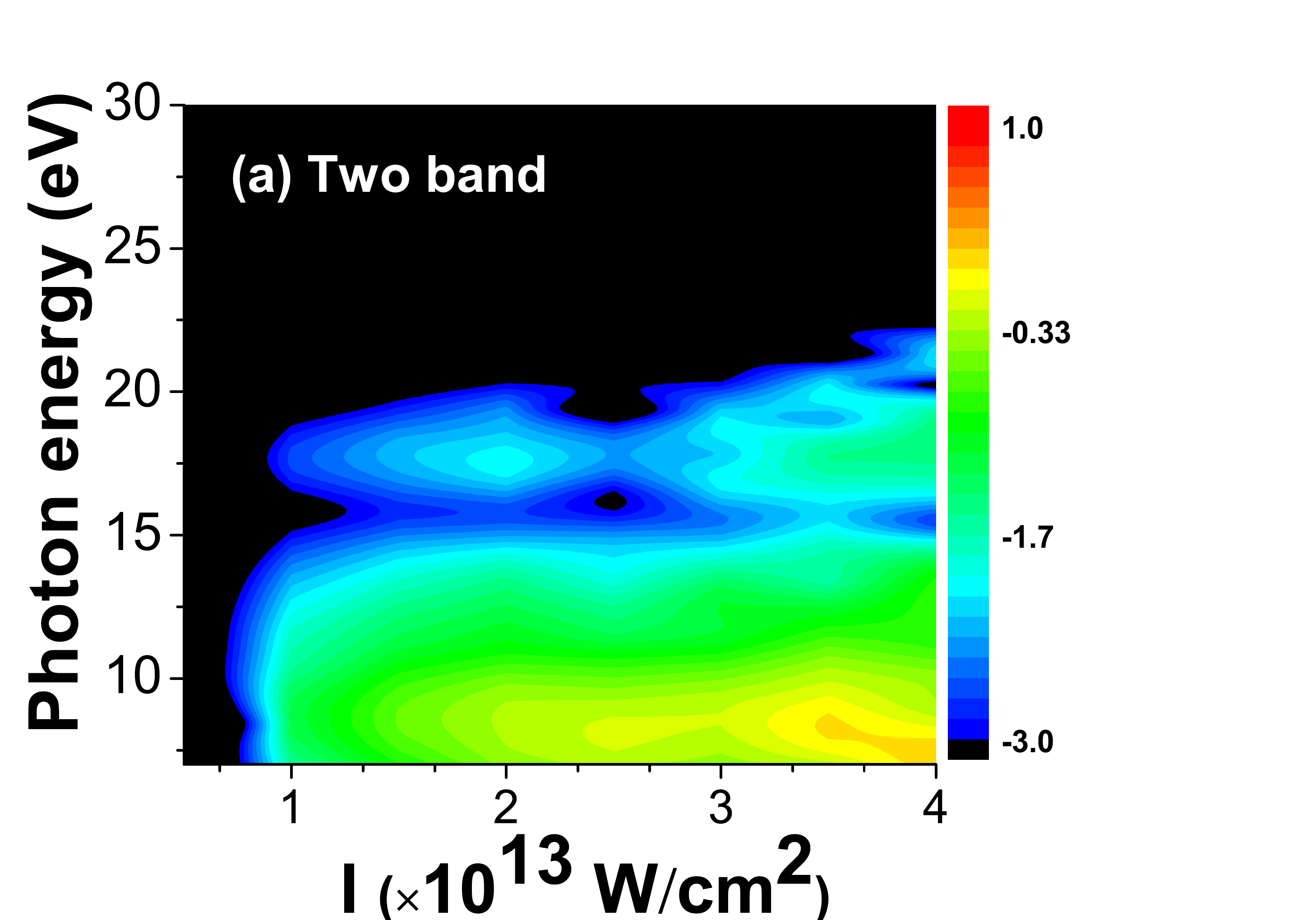}
 		\end{minipage}
 	}
 	\subfigure{
 		\begin{minipage}{4cm}
 			\centering
 			\includegraphics[width=5cm]{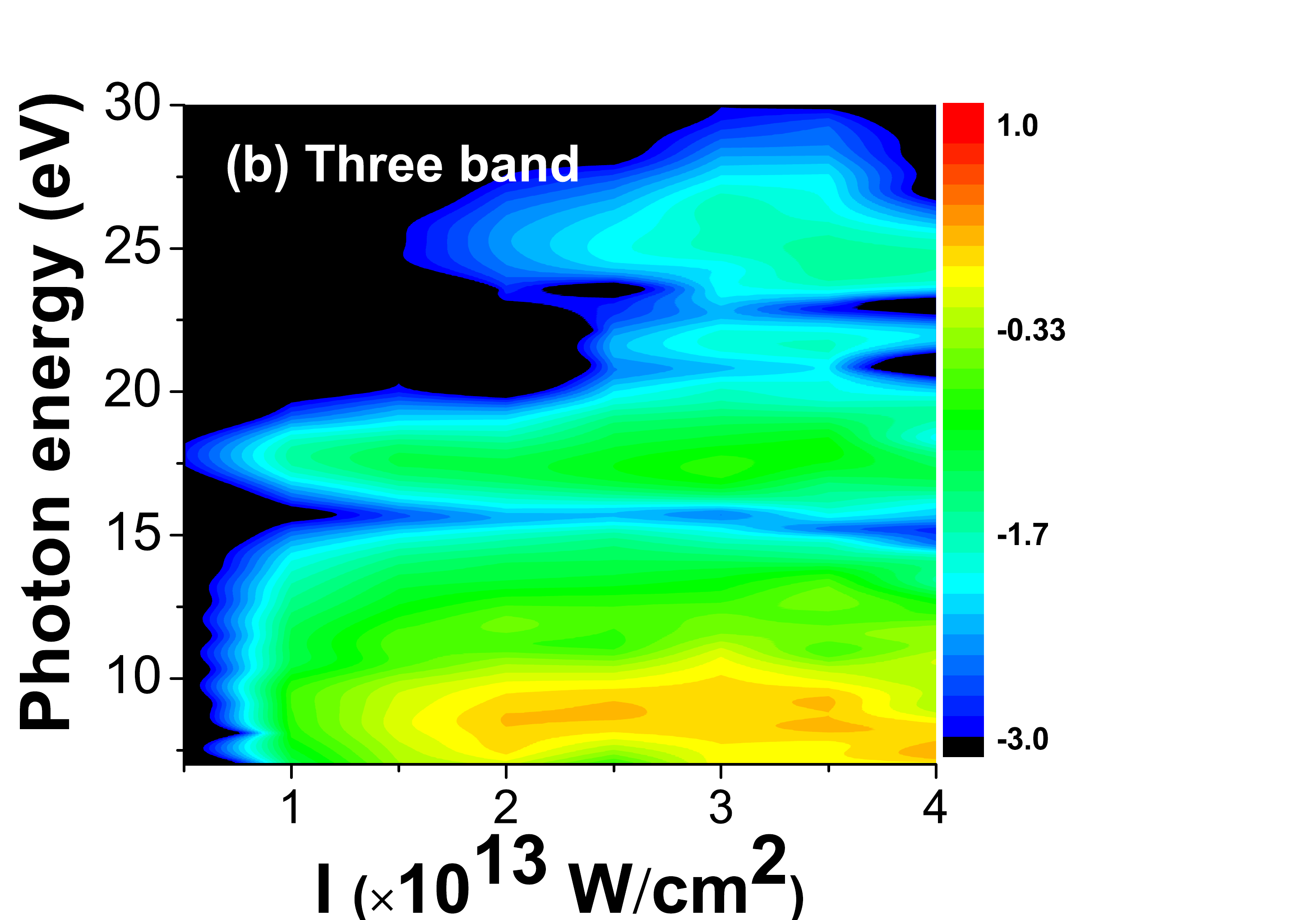}
 		\end{minipage}
 	}
 	\caption{Dependence of crystal HHG spectra from the two-band (a) and three-band (b) models on the driving laser intensity. The duration, wavelength and CEP of the driving laser pulse are 3 fs, 1600 nm, and 0, respectively.}
 \end{figure}
 
 Based on the real TDM from the first-principle calculations, the red solid, black short-dash-dotted and green dash-dotted lines in Fig. 3 present harmonic spectra of MgO crystal in the ultrashort laser pulse, which are generated by the total current, intraband current and the interband polarization, respectively. The laser peak intensity of the incident laser pulse is $3.0\times 10^{13}$ $W/cm^{2}$ (about $1.5$ $V/$\AA), which is lower than the damage threshold of MgO. It is clear that harmonics below/above the bandgap are dominated by intraband/ interband currents, which agree with the recent results for ZnO, MgO, and GaAs in mid-infrared laser pulses \cite{Vampa28,You46,Golde47,Wang48,Wu49}. Figure 3 also shows the harmonic spectrum (blue dashed line) from the TDM based on the first-order $\mathbf{k}\cdot \mathbf{p}$ theory. In this case, intensities of harmonics in the plateau are almost same. However, for the spectrum based on the first-principle calculations, there exists an obvious minimum structure when the photon energy is 16 eV. This distinction in both cases indicates that  the TDM$'$s shape has a significant effect on the crystal harmonic spectra.
 
 \begin{figure} [hbt!]
 	\centering
 	\includegraphics[height=5.5cm,width=8cm]{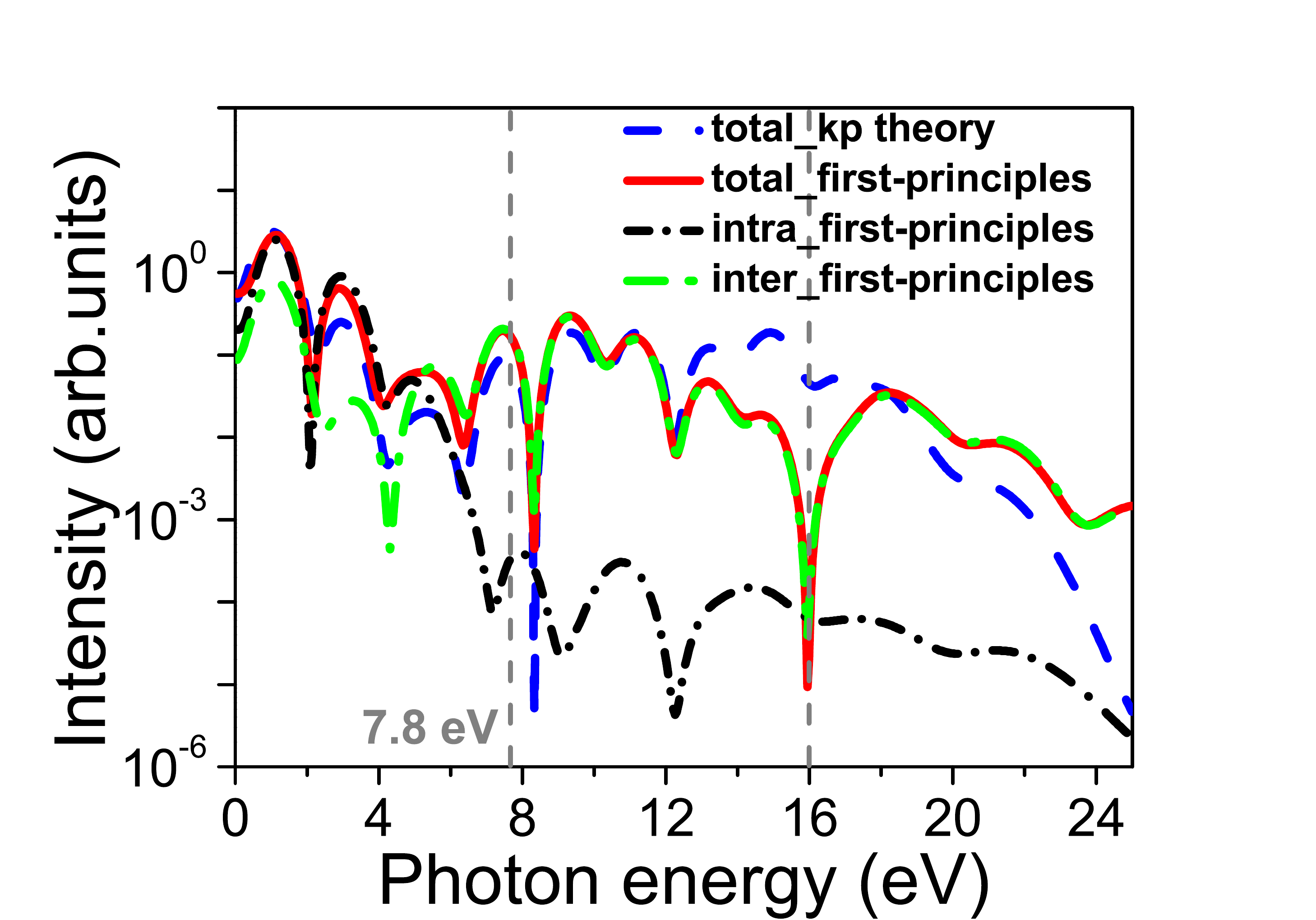}
 	\caption{ Harmonic spectra of MgO from TDMs calculated by first-principle (red solid) and first-order $\mathbf{k}\cdot \mathbf{p}$ theories (blue dashed); the black short dash-dotted and green dash-dotted curves are harmonic spectra produced by intraband and interband currents from TDM with the first-principle calculations, respectively. The laser parameters are the same as in Fig. 2.}
 \end{figure}
 
 Next, we check the influence of laser parameters to the dip structure in harmonic spectra from the real TDM based on the first-principle calculations. Figs. 4(a)-4(d) show the HHG spectra of MgO crystal in laser pulses with different intensity, wavelength, duration, and CEP, respectively. It is found that, the minimum structure at harmonic spectra is almost independent of parameters of the driving laser pulses. Because harmonics in the plateau are mainly originated from the interband polarization, it is natural to deduce that the dip structure is related to the characteristic of the real TDM of MgO crystal. In the following, for intuitively clarifying the minimum feature, we focus on the HHG spectrum from the ultrashort laser pulse, as shown by the red solid line in Fig. 3.
 
 \begin{figure} [hbt!]
 	\centering
 	\includegraphics[height=5.5cm,width=8cm]{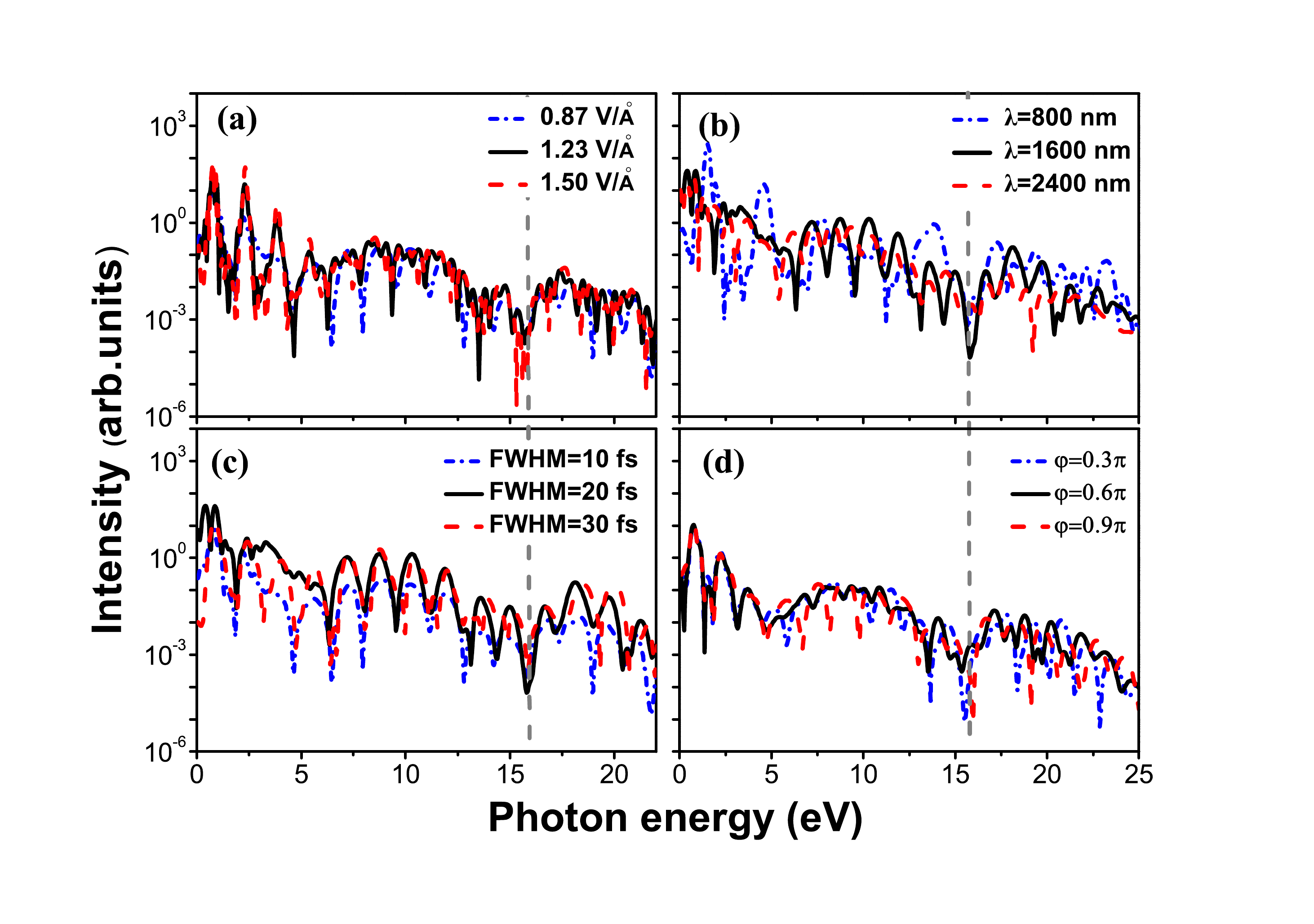}
 	\caption{Harmonic spectra from the real TDM by using first-principle calculations in different laser pulses. (a) the duration, wavelength and CEP are 10 fs, 1600 nm, and 0, respectively; (b) the duration, peak intensity and CEP are 10 fs, $2.0\times 10^{13}$ $W/cm^{2}$, and 0, respectively; (c) the intensity, wavelength and CEP are $2.0\times 10^{13}$ $W/cm^{2}$, 1600 nm, and 0, respectively; (d) the intensity, wavelength and duration are $2.0\times 10^{13}$ $W/cm^{2}$, 1600 nm, and 10 fs, respectively.}
 \end{figure}
 
 In order to further understand the emission process in crystal HHG, time-frequency analyses of the harmonic spectra for first-principle and first-order $\mathbf{k}\cdot \mathbf{p}$ theory cases are presented in Figs. 5(a) and 5(b). In both cases, time-frequency diagrams of HHG are similar, harmonics beyond the bandgap are primarily caused by one quantum path. This result is further confirmed by the harmonic photon energy vs the emission instant calculated from the semiclassical recollision model, as shown in the purple circle curve from Fig. 5. It means that the harmonic photon above the band gap and the crystal momentum at the emission instant has a one-to-one correspondence in the ultrashort laser pulse. Furthermore, the time-frequency distribution in Fig. 5(a) shows one hole at the photon energy with 16 eV, which directly correspond to the dip in the harmonic spectrum (the red solid curve) in Fig. 3. In contrast to the first-principle case, there is no hole at 16 eV in the HHG time-frequency distribution from the first-order $\mathbf{k}\cdot \mathbf{p}$ theory, as shown in Fig. 5(b). The above results further testify that the minimum structure of the harmonic spectrum is closely concerned with the TDM of the crystal.
 
 \begin{figure} [hbt!]
 	\centering
 	\subfigure{
 		\begin{minipage}{4cm}
 			\centering
 			\includegraphics[width=4cm]{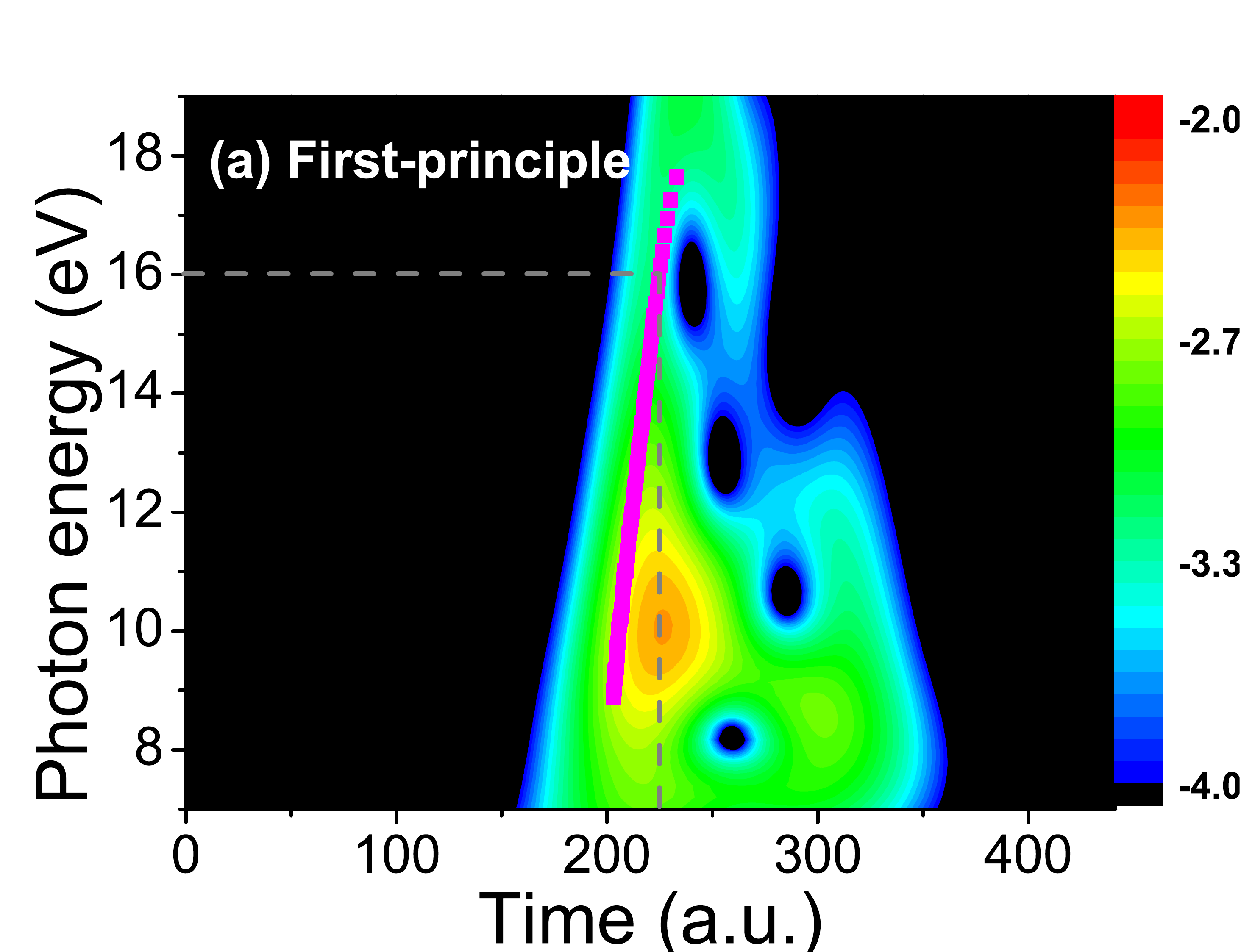}
 		\end{minipage}
 	}
 	\subfigure{
 		\begin{minipage}{4cm}
 			\centering
 			\includegraphics[width=4cm]{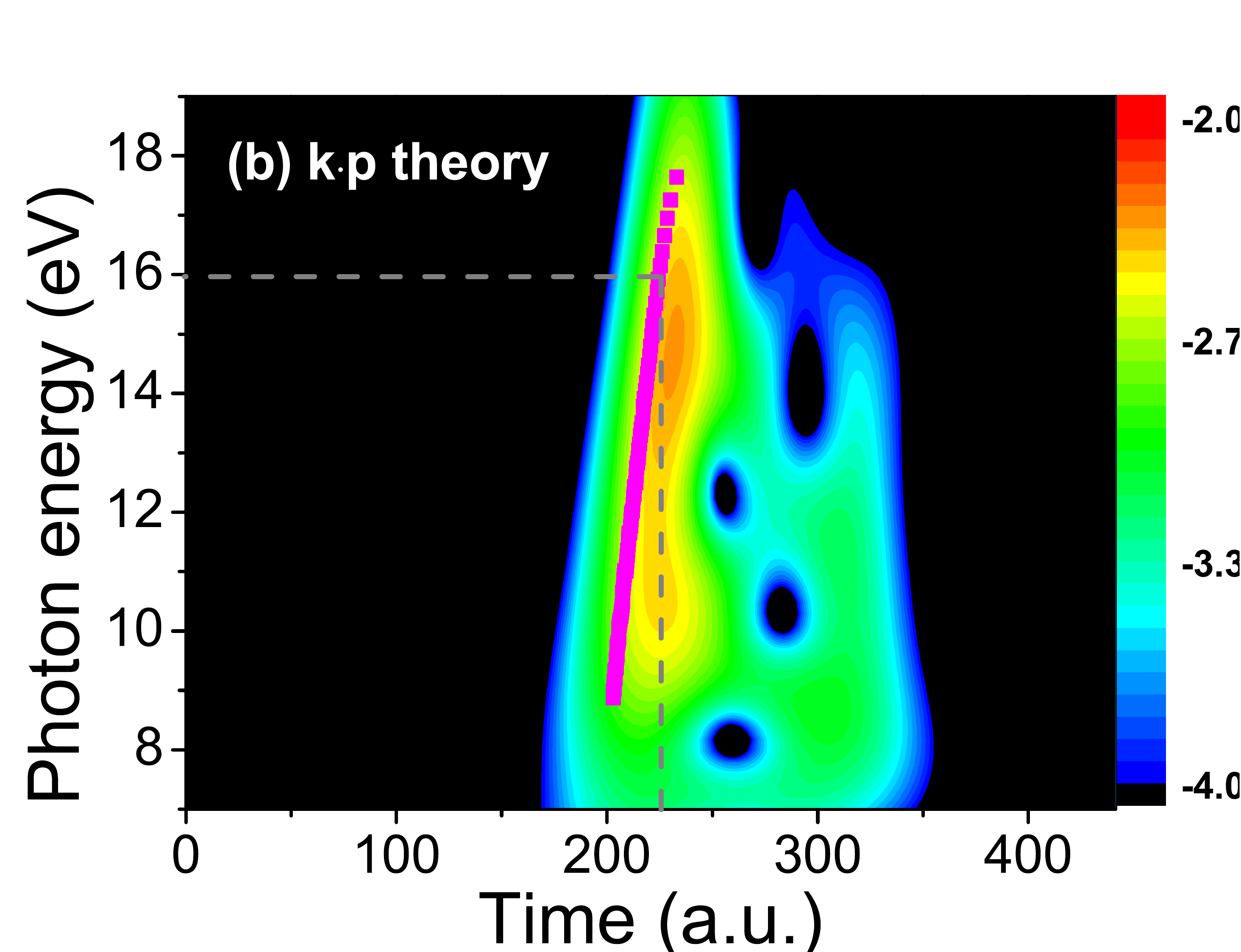}
 		\end{minipage}
 	}
 	\caption{Time-frequency distributions of the HHG corresponding to the red solid (a) and blue dashed (b) curves in Fig. 3. The pink circle curve is the photon energy vs the emission time from the semiclassical recollision model. The laser parameters are the same as in Fig. 2.}
 \end{figure}
 
 The efficiency of HHG from the interband current is proportional to the population of the electron (hole) and the TDM between conduction and valence bands at the recombination time $t_{r}$. It can be observed that, the emission instant is 225 a.u. from Fig. 5 when the energy of the harmonic photon is equal to 16 eV. To clearly address the physics of the dip structure in harmonic spectra, we examine populations of electrons in the conduction band at this emission instant. Figure 6 shows electronic populations of the conduction band for the first-principle and first-order $\mathbf{k}\cdot \mathbf{p}$ cases, respectively. From Figs. 1(a) and 1(b), one can find that the bandgap between two bands is exactly equal to 16 eV for the crystal momentum at k=$\pm$0.6 $\pi/a$. In Fig. 6, populations at these crystal momentums for the emission$'$s instant 225 a.u. are marked by the cross of dashed lines. For both cases, populations at k=$\pm$0.6 $\pi/a$ with  $t_{r}$=225 a.u. are no essential difference. This means that the dip structure of the HHG spectrum is almost independent of the electronic population at the recombination instant.
 
 \begin{figure} [hbt!]
 	\centering
 	\includegraphics[height=5.5cm,width=8cm]{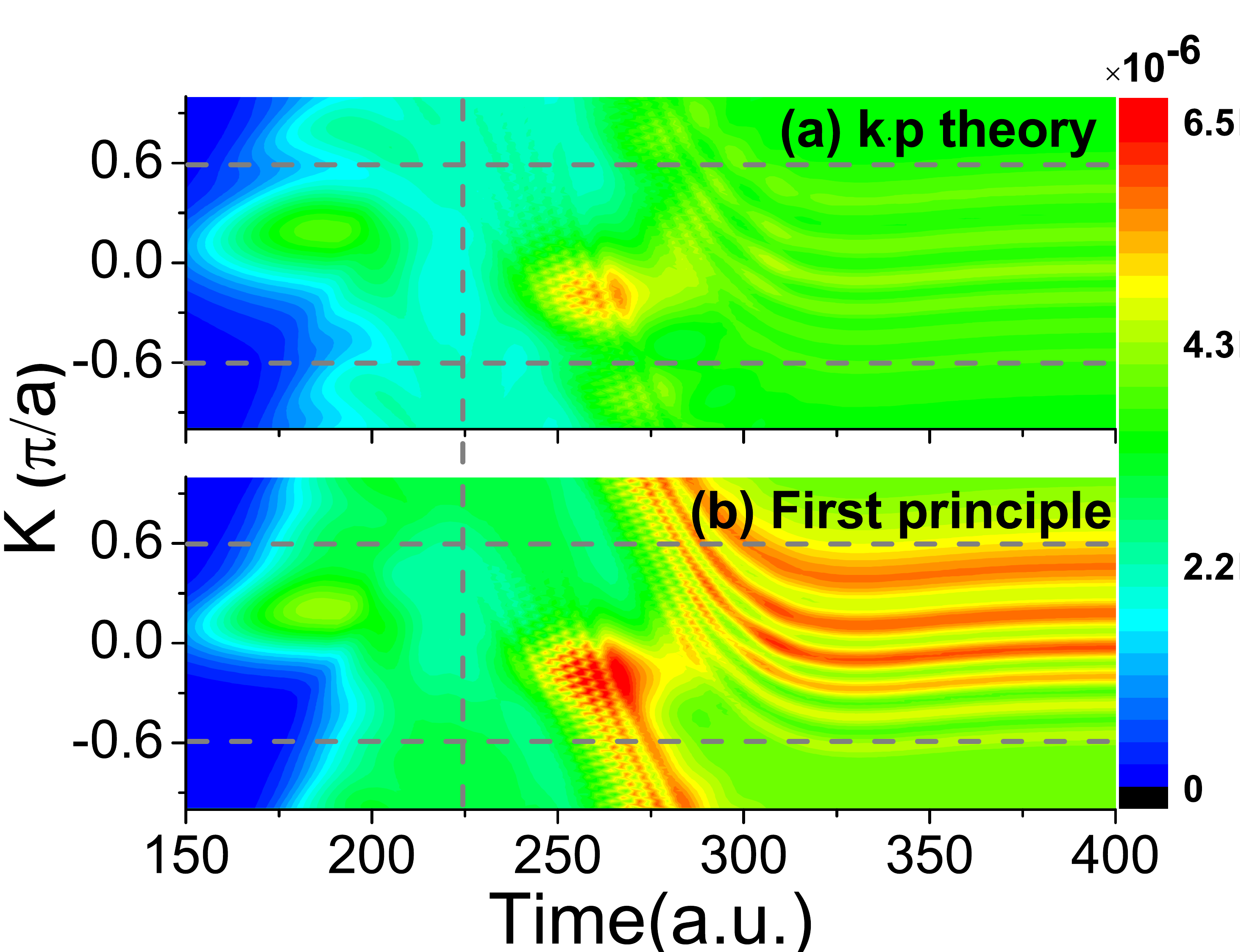}
 	\caption{Electronic populations of the conduction  bands for the dipole moments from first first-order $\mathbf{k}\cdot \mathbf{p}$  (a) and first-principle (b) calculations, respectively. }
 \end{figure}
 
 Now that we know the dip structure of the harmonic spectrum is related to the k-dependent TDM, the contribution from different crystal momentums to harmonics above the bandgap should be analyzed. Figs. 7(a) and 7(b) provide a comparison between distributions of harmonic intensities at different k from the first first-order $\mathbf{k}\cdot \mathbf{p}$ theory and first-principle calculations, respectively. Here, in order to distinctly reveal distributions of harmonics intensities, we focus on harmonics produced at the main emission times from 200 a.u. to 250 a.u.. In the case of the real TDM from the first-principle calculations, the amplitude value of TDM near k=$\pm$0.6 $\pi/a$ is close to zero as shown by the orange solid curve in Fig. 7(b), which results in no distribution of the harmonic intensity , as presented in Fig. 7(b). In the first-order $\mathbf{k}\cdot \mathbf{p}$ theory case, TDM has bigger values near k=$\pm$0.6 $\pi/a$, which induces a clear distribution of the harmonic intensity, as observed from Fig. 7(a). Above all, it can be demonstrated that the valley shape of the TDM from the first-principle calculations results in the dip structure in the harmonic spectra.
 
 \begin{figure} [hbt!]
 	\centering
 	\subfigure{
 		\begin{minipage}{4cm}
 			\centering
 			\includegraphics[width=4cm]{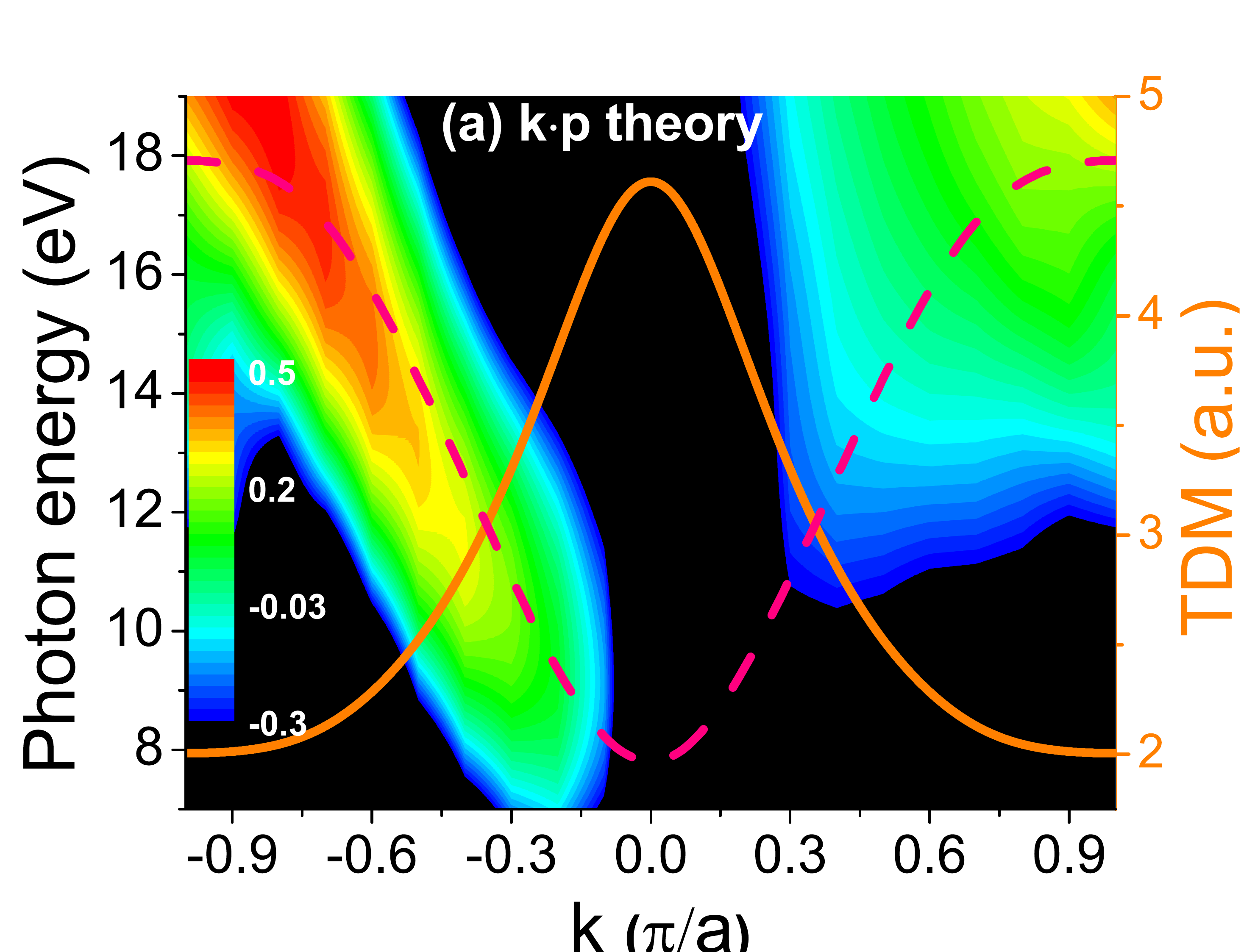}
 		\end{minipage}
 	}
 	\subfigure{
 		\begin{minipage}{4cm}
 			\centering
 			\includegraphics[width=4cm]{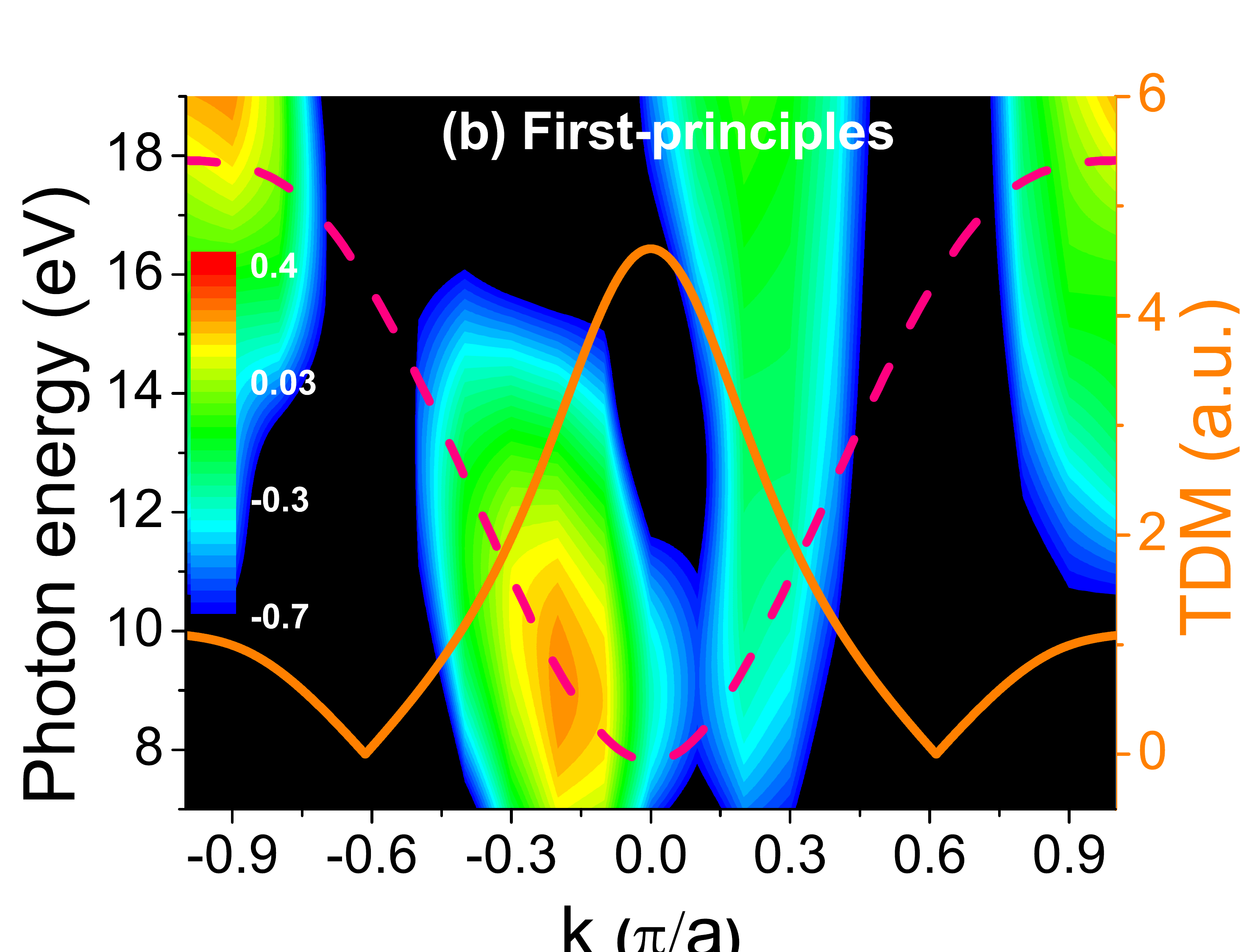}
 		\end{minipage}
 	}
 	\caption{ The contribution of different crystal momentums to harmonics based on the first first-order $\mathbf{k}\cdot \mathbf{p}$ (a) and first-principle calculations (b), respectively. The orange solid curves are the TDMs from the two cases, and the pink dashed curve is the bandgap between two bands. The laser parameters are the same as in Fig. 2.}
 \end{figure}
 
 Finally, we explore dependences of amplitudes of TDMs from the first-principle calculations and the first-order $\mathbf{k}\cdot \mathbf{p}$ theory with the bandgap, as shown by the green short-dotted and black dotted curves in Fig. 8. The corresponding harmonic spectra in both cases are also presented in Fig. 8. In the case of the first-order $\mathbf{k}\cdot \mathbf{p}$ theory, TDM and the harmonic spectrum near 16 eV have no minimum structure. However, for the case of the first-principle calculations, the TDM$'$s valley at 16 eV directly coincides with the minimum of the harmonic spectrum. Thereby, we can draw a conclusion that, because the amplitude of TDM between valence and conduction bands exists zero values, harmonic spectra from MgO crystal also exhibit the Cooper minimum structure, which is similar to harmonic spectra from gaseous media. 
 
 \begin{figure} [hbt!]
 	\centering
 	\includegraphics[height=5.5cm,width=8cm]{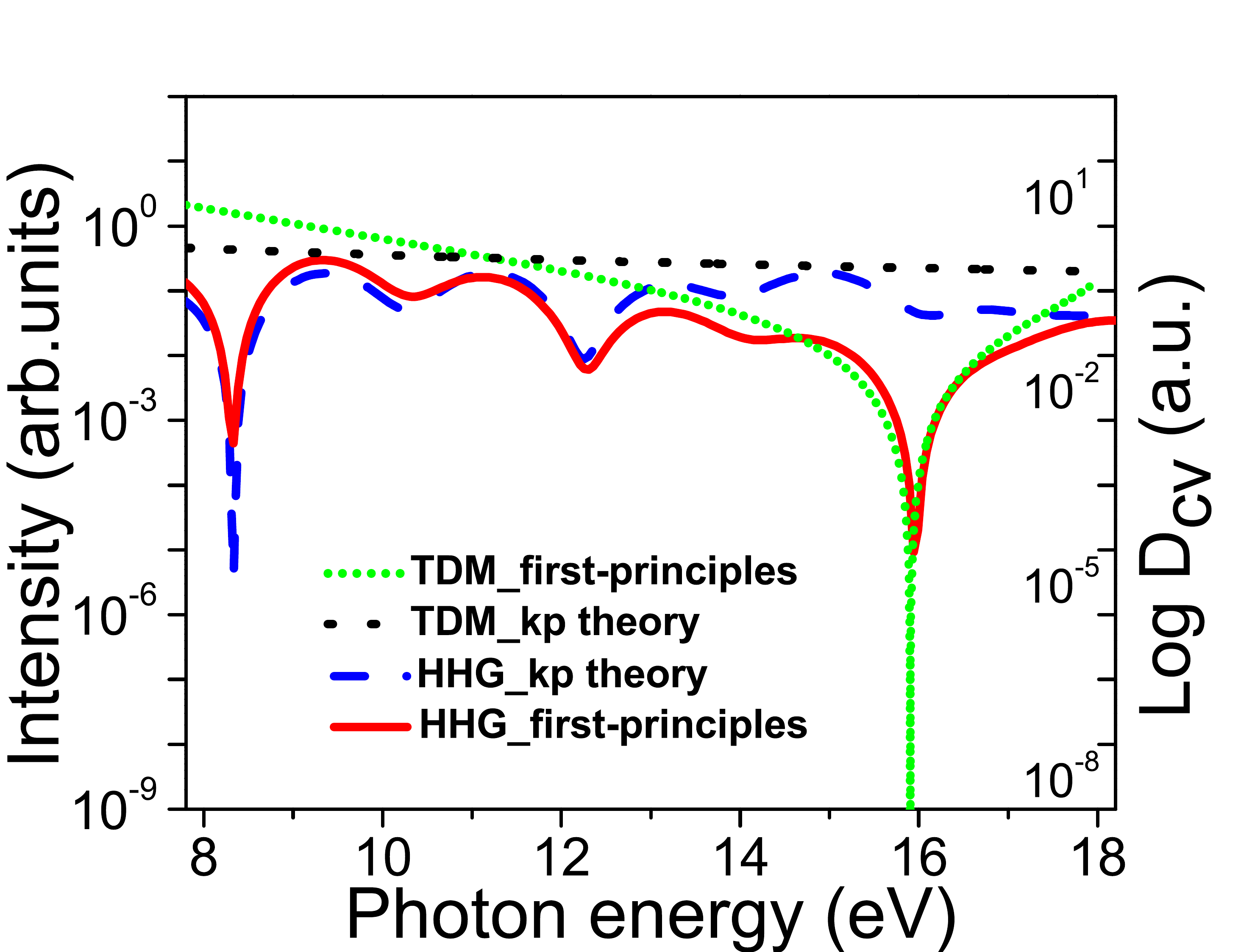}
 	\caption{Values of TDMs from the first-principle calculations  (green short-dotted line) and the first-order $\mathbf{k}\cdot \mathbf{p}$ theory (black dotted line) versus the bandgap, and the corresponding harmonic spectra shown by red solid and blue dashed curves in Fig. 3. }
 \end{figure}

\section{SUMMARY}

In conclusion, we have demonstrated that harmonic spectra of MgO crystal in the ultrashort laser pulse have the Cooper minimum structure. By comparing harmonic spectra from TDMs of the first-order $\mathbf{k}\cdot\mathbf{p}$ theory and the first-principle calculations, it is confirmed that, the shape of TDM plays an important role in the generation of the HHG spectrum, and the valley of the real TDM from the first-principle calculations lead to the dip structure near 16 eV in the harmonic spectra from MgO crystal. More importantly, by taking the valley-dip correspondence as the benchmark, the emitted photon energy and the crystal momentum have a one-to-one match, and the intensity of the harmonic from an ultrashort laser pulse is approximately proportional to the square of the TDM's value. Thereby, the k-dependent bandgap and TDM between valence and conduction bands are hoped to be mapped by harmonics with energies above the minimum bandgap, which will pave a new way to the all-optical reconstruction of the electronic band structure by taking advantage of the crystal HHG.

\section*{ACKNOWLEDGEMENT}
The authors sincerely thank Prof. Ruifeng Lu for providing the code. J. G. Chen is supported by the National Natural Science Foundation of China under Grant No. 11975012. Project supported by the National Key R\&D Program of China (Grant No. 2017YFA0403300), the National Natural Science Foundation of China (Grant Nos. 11774129, 11627807), the Jilin Provincial Research Foundation for Basic Research, China (Grant No. 20170101153JC) and the Science and Technology project of the Jilin Provincial Education Department (Grant No. JJKH20190183KJ).

\end{document}